\documentclass[12pt]{article}
\usepackage[dvips]{color}
\usepackage{epsfig}
\usepackage{amsmath}
\usepackage{graphicx}

\begin{document}

\title{\bf The Thermodynamic Efficiency in Static and Dynamic Black Holes}
\author{{Kh. Jafarzade $^{a}$ \thanks{Email: kh.Jafarzadeh@stu.umz.ac.ir} \hspace{1mm},
J. Sadeghi $^{a}$\thanks{Email:
pouriya@ipm.ir}} \\
$^a${\small {\em Sciences Faculty, Department of Physics,University
of Mazandaran ,}}\\{\small {\em P .O .Box 47415-416, Babolsar,
Iran}}} \maketitle

\begin{abstract}
We note that, in  extended phase space the cosmological constant is
a thermodynamic variable. In this paper, this cosmological constant
lead us to consider a black hole as a heat engine. So, here we take
advantage from holographic heat engine and study two kind of
different black holes. We first investigate a static black hole
(Dyonic BH) and consider the necessary condition to have high
efficiency.  Also we continue our investigation for dynamic black
hole (rotating charged black hole) and study the effect of rotating
parameter on the thermodynamic efficiency of holographic heat
engine. We show that the rotating parameter has a more effective
role than electric charge in thermodynamic efficiency.

\end{abstract}
\section{Introduction}
first time the thermodynamics of black holes was introduced by
Hawking, Carter and Bardeen. They believed that black holes are
thermodynamics and their temperature is related to event horizon
[1]. It means that dynamic laws of black holes are similar to
thermodynamic laws [2]. The similarity of four mechanic laws of
black holes and thermodynamic laws formulated by Hawking, Carter,
Bardeen and Chistodoulou [3-4]. In classical view of point,
parameters of black hole as  mass $M$, surface gravity $\kappa$ and
area $A$ relate to the energy $U$, temperature $T$ and entropy $S$
of thermodynamic system which are given by [5-8],
\begin{eqnarray}
M=U~~~~,~~~~   T=\frac{\kappa}{2\pi}~~~~,~~~~S=\frac{A}{4},
\end{eqnarray}
and this subject has been extended to include the pressure $P$ and
volume $V$ of black hole [9-20]. The cosmological constant of space
time relates to pressure as
 $P=-\frac{\Lambda}{8\pi}$ and thermodynamic volume of black holes is
 defined as: $V=(\frac{\partial M}{\partial P})_{S,\phi_i,J_k}$ (here we are using geometric units $G_N= \hbar = c = k = 1$). The black holes may have other parameters such as gauge charge $q_i$ and angular momentum $J_j$ with their conjugate, as potential $\phi_i$ and angular velocity $\Omega_j$ respectively.\\
 Recently, the variation of cosmological
constant $\Lambda$, in the first law of black holes thermodynamics
has been attention by Ref.s. [21]. This lead us to consider the
pressure as  a thermodynamic variable, where the  mass will be
enthalpy [14,22],
\begin{equation}
M=H\equiv U + PV.
\end{equation}
The first law of black holes thermodynamics in an extended
 phase space in four dimensions with an
electric charge and rotation is,
\begin{equation}
dM = TdS + VdP + \Phi dq + \Omega dJ,
\end{equation}
when $P$ is treated as constant (cosmological constant is not
allowed to vary), the above relation reduces to the standard first
law in the "non-extended" phase space. The thermodynamic volume of
static black holes is equal to volume of horizon radius. For example
in 4-dimensional Reissner- Nordstrom
 and  Schwarzchild black holes we have following equation,
\begin{equation}
V=\frac{4}{3}\pi r_H^3,
\end{equation}
where the entropy is related to horizon radius and volume in static
black holes. Thermodynamics of black holes not only determines
standard thermodynamic variables such as temperature and entropy but
also has extensive phase structure in analogy with known non
gravitational thermodynamic systems and also it admit critical
phenomena.
 Historically, the study of
thermodynamic  properties of $AdS$ black hole was started by
impressive article of Hawking and Page [23], they have shown a phase
transition in phase space of $AdS$ Schwarzchild black hole (non
rotating uncharged). Then papers about phase transition and critical
phenomena of systems with more complicated background increased
[24,25]. The critical behavior of $RN-AdS$ black hole in non
extended phase space ($\Lambda$=cte and  P=cte), at canonical
ensemble (fixed charge) was studied in [26,27] that was observed
phase transition behavior is similar to Liquid-gas phase transition.
Also, the critical behavior of this black hole in extended phase
space is remarkably coincidence with Van der Waals fluid studied by
[28]. In Ref. [29], authors considered a Dyonic black hole in (3+1)
dimensions (this black hole has a magnetic charge in additional to
electric charge) and they have shown that putting black hole into
ensemble of fixed electric charge and magnetic charge the obtained
results is similar to [25-27] except the charge is redefined by
$q_E^2\rightarrow q_E^2 +q_M^2$. The quantum behavior of this black
hole was studied in Ref.[30]. In quantum approach, the effect of
thermal fluctuations is considered. Thermal fluctuations arise due
to quantum fluctuations in the geometry of space-time and they
appear in the black hole entropy as logarithmic term [31-34]. But
it's not our purpose in this paper. Here, we study thermodynamic
black hole from the classical point of view. After studying of phase
transition, it is interesting to define classical cycles for black
holes like usual thermodynamic systems. It means that when   small
black hole translate to large black hole we need again the large
black hole translate   to small black hole, in other words the
system must be back  to primary state. On the other hand, the
holographic heat engine has been studied  for charge $AdS$ black
hole [35]. This paper and above information give us motivation to
study heat engine in dynamic black hole and compare with static
black hole. In this paper first of all we are going to consider
Static and Dynamic black holes and review some thermodynamic
properties as thermodynamic cycles and heat engines generally. These
thermodynamic cycles and heat engines help us to arrange black hole
as Carnot cycle and obtain the corresponding efficiency for the
static and dynamic black holes.
\section{Thermodynamic cycles and heat engines}
By using volume, pressure, temperature and entropy we can calculate
heat energy and mechanical useful work. We start with equation of
state (function of $P(V,T)$) and define an engine as a close path in
$P-V$ plane which receives $Q_H$ and gives $Q_C$. From the first law
of thermodynamics, total mechanical work is defined as, $W=Q_H -
Q_C$. Therefore, the efficiency of heat engine is,
$\eta=\frac{W}{Q_H}=1-\frac{Q_C}{Q_H}$.\\
Some of the classic cycles involve a pair of isotherms at
temperature $T_H$ and $T_C$ ( $T_H>T_C$)  where there are isothermal
expansion and  compression while some heat is absorbed and some heat
is exited respectively. By using different methods one can relate
two systems with each other. The first method here is isochoric path
like classical Stirling cycle and the
second one is adiabatic path like classical Carnot cycle.\\
We know that a whole heat engine is fully reversible (since the
total entropy flow zero). Therefore, we note here the maximum
efficiency  will be  Carnot efficiency $(\eta=1-\frac{T_C}{T_H})$ .
Any higher efficiency would violate the second Law. It is very
important that how can reach to such efficiency in heat engine of
black holes to preserve the second law. So, the form of path for the
definition of cycle is important. As we know, in the static black
holes the thermodynamic volume $V $ and  entropy $S$  are not
independent. It means that adiabats and isochores are the same, in
that case Carnot and Stirling coincide to each other. So, the
efficiency of cycle can be calculated easily.

\begin{figure}
\hspace*{1cm}
\begin{center}
\epsfig{file=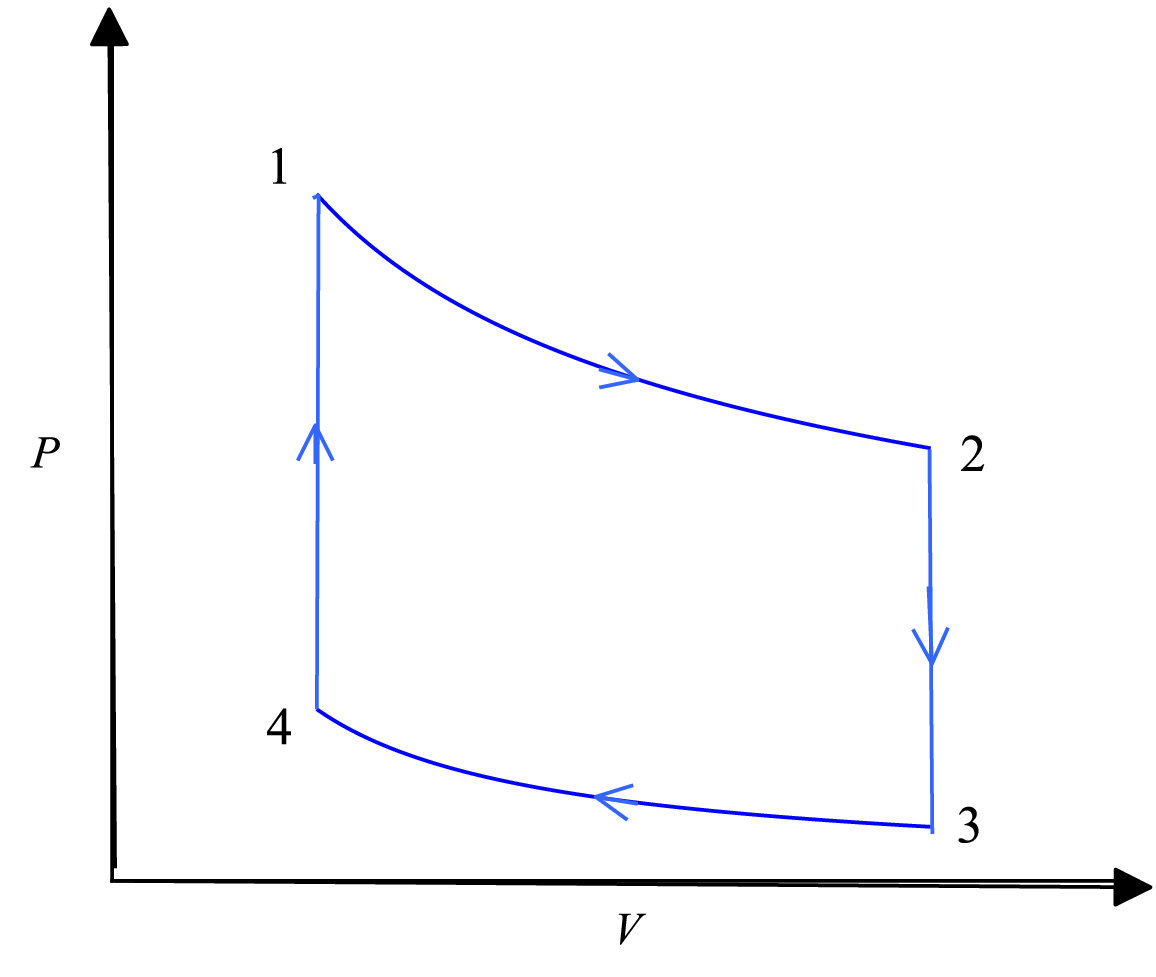,width=7cm} \caption{\small{The Carnot engine.}}
\end{center}
\end{figure}

So along the upper isotherm, we have the following heat flow,
\begin{equation}
Q_H =T_H\Delta
S_{1\rightarrow2}=T_H\bigg(\frac{3}{4\pi}\bigg)^\frac{2}{3}\pi(V_2^{\frac{2}{3}}-V_1^{\frac{2}{3}}),
\end{equation}
and also along the lower isotherm the heat flow will be as,
\begin{equation}
Q_C =T_C\Delta
S_{3\rightarrow4}=T_C\bigg(\frac{3}{4\pi}\bigg)^\frac{2}{3}\pi(V_3^{\frac{2}{3}}-V_4^{\frac{2}{3}}).
\end{equation}
Since $V_1=V_4$ and $V_2=V_3$, the efficiency becomes,
\begin{equation}
\eta=1-\frac{T_C}{T_H}.
\end{equation}
So, we take advantage from above thermodynamic cycles and heat engines information and study the static
and dynamic black hole.
\section{Thermodynamical properties of static black hole and heat engine}
Now we are going to consider the static black hole system (dyonic
black hole). The solution of dyonic-dilaton $AdS$ black hole can be
used for maximal gauge supergravity in 4-dimension [36]. In that
case the corresponding action will be as,
\begin{equation}
I=\frac{1}{16\pi G_4}\int d^4 x\sqrt{g}(-R + F^2 -\frac{6}{b^2}).
\end{equation}
The equation of motion is given by,
\begin{eqnarray}
R_{\mu\nu}-\frac{1}{2}g_{\mu\nu}R
-\frac{3}{b^2}g_{\mu\nu}=2(F_{\mu\lambda}F_\nu^\lambda
-\frac{1}{4}g_{\mu\nu}F_{\alpha\beta}F^{\alpha\beta});\nonumber \\&&
\hspace{-71mm}
                  \nabla_\mu F^{\mu\nu}=0.
\end{eqnarray}
One can write a static spherically symmetric solution to this
action which is given by,
\begin{equation}
ds^2=-f(r)dt^2 +\frac{1}{f(r)}dr^2 + r^2d\theta^2 + r^2\sin^2\theta
d\phi^2,
\end{equation}
where
\begin{equation}
f(r)=(1 + \frac{r^2}{b^2} -\frac{2M}{r}+\frac{q_E^2+q_M^2}{r^2}).
\end{equation}
The electromagnetic four-potential $A_\mu$ is,
\begin{equation}
A=(-\frac{q_E}{r}+\frac{q_E}{r_+})dt + (q_M\cos\theta)d\phi ,
\end{equation}
where $q_E$ , $q_M$ and $M$ are electric, magnetic charge and
mass of black hole respectively. The horizon of black hole is given
by,
\begin{equation}
f(r_+)=(1 + \frac{r_+^2}{b^2}
-\frac{2M}{r_+}+\frac{q_E^2+q_M^2}{r_+^2})=0,
\end{equation}
and electric potential $\Phi_E$ is defined by following equation,
\begin{equation}
\Phi_E=\frac{q_E}{r_+}.
\end{equation}
The Hawking temperature of this black hole is  following,
\begin{equation}
T=\frac{1}{\beta}=\frac{1}{4\pi r_+}[1+\frac{3r_+^2}{b^2} - \Phi_E^2
-\frac{q_M^2}{r_+^2}].
\end{equation}
The pressure is given by [28,37],
\begin{equation}
P=-\frac{\Lambda}{8\pi}=\frac{3}{8\pi}\frac{1}{b^2},
\end{equation}
and the thermodynamical volume is,
\begin{equation}
V=\frac{4\pi}{3}r_+^3.
\end{equation}
We can write the temperature in equation (15) in terms of $S$ and $P$ as
follows,
\begin{equation}
T=\frac{1}{4\sqrt{\pi}}\frac{1}{\sqrt{S}}(1 +8PS - \Phi_E^2
-\frac{\pi q_M^2}{S}).
\end{equation}
The equation of state is given by,
\begin{equation}
P=\frac{T}{\upsilon}-\frac{1-\Phi_E^2}{2\pi
\upsilon^2}+\frac{2q_M^2}{\pi \upsilon^4},
\end{equation}
where $\upsilon=2r_+$ can be identified with the specific volume of
the system [29]. This equation describes different phases of a
dyonic black hole in a fixed electric potential and magnetic charge
ensemble which is similar to extended liquid-gas phase diagram.  In
thermodynamic,  the heat capacity is an important measurable
physical quantity. It determine the amount of requisite heat to
change the
 temperature of an object by a given amount. There are two different
heat capacities for a system, heat capacity at constant pressure and
heat capacity at constant volume. Heat capacity can be calculated by
the standard thermodynamic relations, which is given by,
\begin{equation}
C_V=T\frac{\partial S}{\partial T}\bigg|_V ~~~~~, ~~~~~
C_P=T\frac{\partial S}{\partial T}\bigg|_P.
\end{equation}
The entropy is expressed as,
\begin{equation}
S=\frac{A_H}{4}=\pi
r_+^2=\pi\bigg(\frac{3V}{4\pi}\bigg)^\frac{2}{3}.
\end{equation}
By using Eq. (20) and Eq. (21) we have $C_V=0$ and $C_P$ which is
given by,
\begin{equation}
C_P=2S\frac{(8PS^2 + S(1-\Phi_E^2)-\pi q_M^2)}{(8PS^2 +
S(1-\Phi_E^2)+3\pi q_M^2)}~.
\end{equation}
One can write $P$ from equation (19) as a function of thermodynamical volume, so we have,
\begin{equation}
P=\frac{T}{V^\frac{1}{3}}\bigg(\frac{\pi}{6}\bigg)^\frac{1}{3}-\frac{1-\Phi_E^2}{2\pi
V^\frac{2}{3}}\bigg(\frac{\pi}{6}\bigg)^\frac{2}{3}+\frac{2q_M^2}{\pi
V^\frac{4}{3}}\bigg(\frac{\pi}{6}\bigg)^\frac{4}{3}.
\end{equation}
In $P-V$ diagram we fixed $(T,~ \Phi_E,~ q_M)$, so  in that case we
draw $P$ with respect $V$ as  two figures $2a$ and $2b$ .
\begin{figure}
\hspace*{1cm}
\begin{center}
\epsfig{file=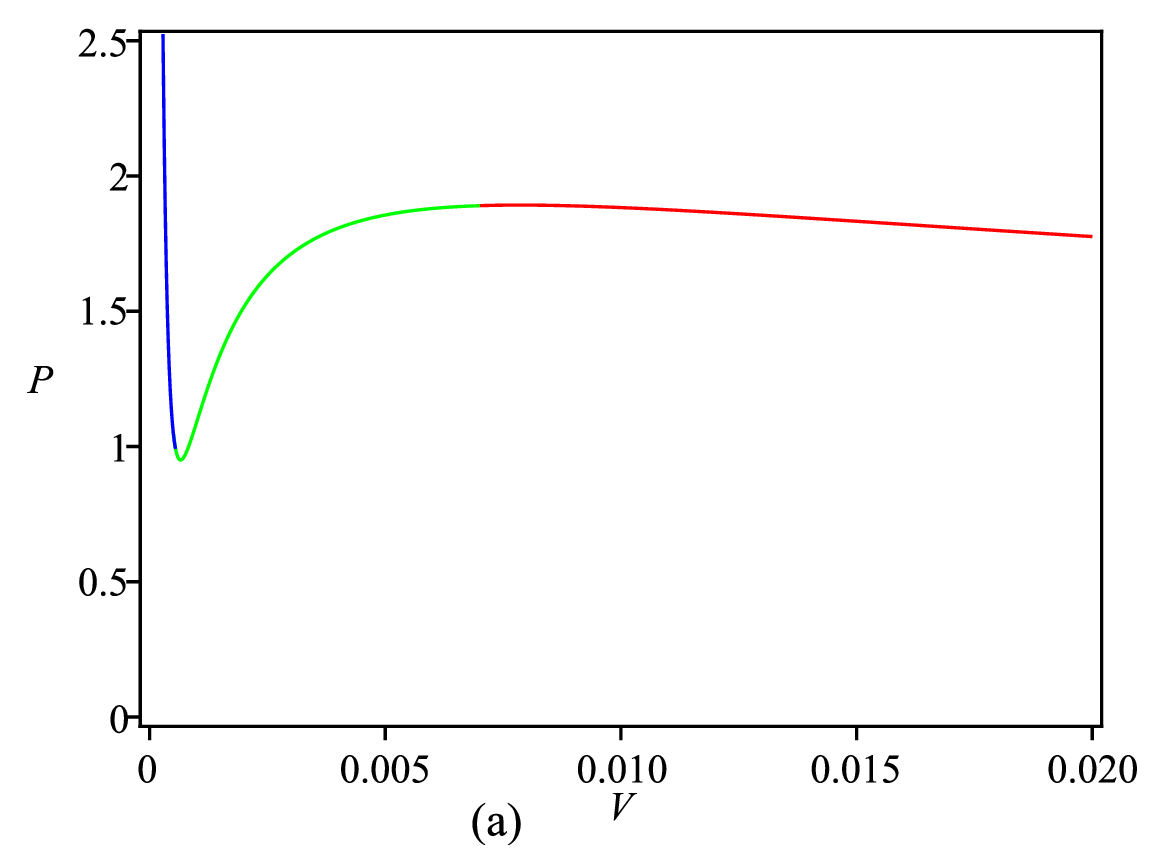,width=7cm}
\epsfig{file=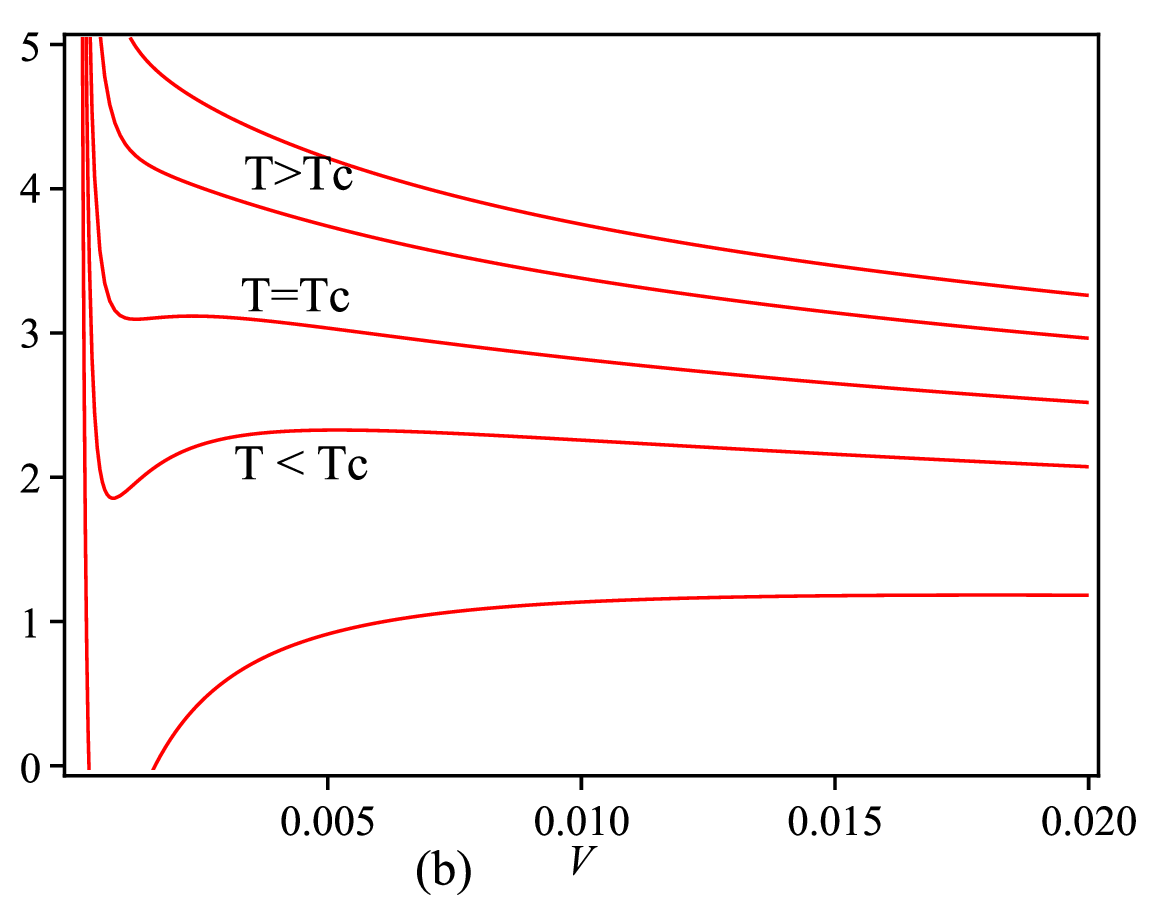,width=7cm}\caption{\small{(a) P-V diagram for all
other parameters held constant ($q_M=0.028$, $\Phi_E=0.35$,
$T=1.05$). (b) P-V diagram for fixed ($q_M=0.028$, $\Phi_E=0.35$)
and varying T.}}
\end{center}
\end{figure}
In figure (2a), there are three different solutions for static black
hole horizon. These are identified by  branch-1 (blue), branch-2
(green) and branch-3 (red). The small black hole (SBH) just exist
for the large $P$, so we have branch-1. For the low $P$, we only
have the large black hole (LBH) as branch-3. On the other hand , the
branch-2 indicates  thermodynamical unstable phase. Now we back to
figure (2b) and  see that the branch-2 is disappeared by the
condition $(T>T_{c})$ and phase transition is happened for $T<T_C$.
In phase transition, SBH reduces to LBH. If  LBH transfers to SBH,
we can define classical cycle for such black hole. The SBH absorbs
heat $Q_H$ along isothermal expansion and exists at high pressure.
Actually, an explicit expression $C_P$ lead us to have a new engine
which include  two isobars and two isochores/adiabatas as figure(3).
The work done in this cycle is,

\begin{figure}
\hspace*{1cm}
\begin{center}
\epsfig{file=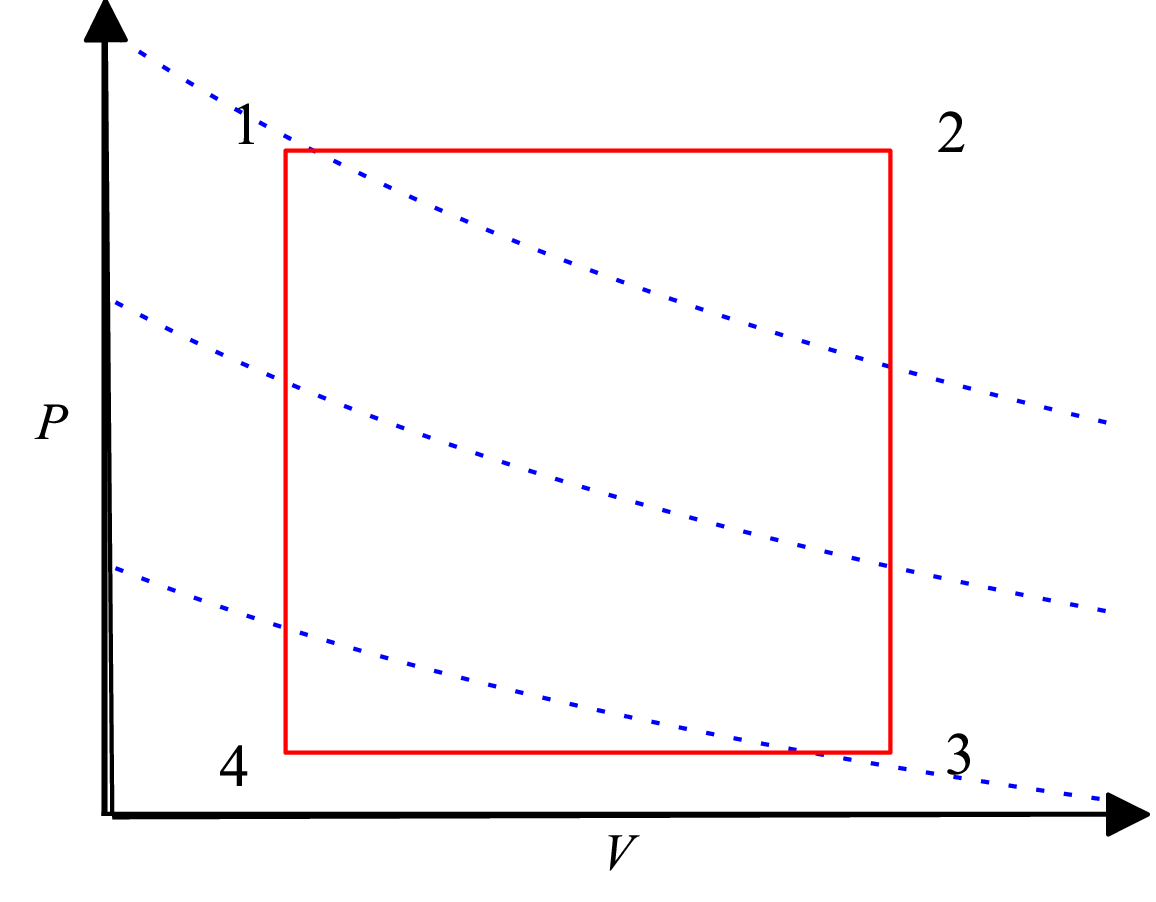,width=7cm} \caption{\small{Our other engine.}}
\end{center}
\end{figure}

\begin{eqnarray}
W=\oint PdV\nonumber \\&& \hspace{-31mm} W_{total}=W_{1\rightarrow2}
+W_{3\rightarrow4}=P_1(V_2-V_1)+P_4(V_4-V_3)\nonumber \\&&
\hspace{-31mm}
W_{total}=\frac{4}{3\sqrt{\pi}}(P_1-P_4)(S_2^\frac{3}{2}-
S_1^\frac{3}{2}).
\end{eqnarray}
The upper isobar will give us following  heat  $Q_H$,
\begin{equation}
Q_H=\int_{T_1}^{T_2} C_P(P_1,T)dT.
\end{equation}
The corresponding entropy and $C_P$ in high pressure and temperature are given by,
\begin{eqnarray}
S=\frac{\pi
T^{2}}{4P^{2}}-\frac{(1-\Phi_{E}^{2})}{4P}-\frac{(1-\Phi_{E}^{2})^{2}}{16\pi
T^{2} }+...\nonumber \\&& \hspace{-80mm}C_{P}=\frac{\pi
T^{2}}{2P^{2}}-\frac{(1-\Phi_{E}^{2})}{2P}-\frac{(1-\Phi_{E}^{2})^{2}}{8\pi
T^{2} }+...~,
\end{eqnarray}
and
\begin{equation}
Q_H=\frac{4P_{1}}{3\sqrt{\pi}}(S_{2}^{\frac{3}{2}}-S_{1}^{\frac{3}{2}})-\frac{(1-\Phi_{E}^{2})}{\sqrt{\pi}}(S_{2}^{\frac{1}{2}}-S_{1}^{\frac{1}{2}})+\frac{(1-\Phi_{E}^{2})^{2}}{16\sqrt{\pi}P_{1}}(S_{2}^{-\frac{1}{2}}-S_{1}^{-\frac{1}{2}}).
\end{equation}
The above equation help us to arrange the efficiency for the static black hole which is obtained by,
\begin{equation}
\eta=\bigg(1 - \frac{T_C}{T_H}\bigg)\bigg\{
1-\frac{3(\Phi_{E}^{2}-1)(S_{2}^{\frac{1}{2}}-S_{1}^{\frac{1}{2}})}{4P_{1}(S_{2}^{\frac{3}{2}}-S_{1}^{\frac{3}{2}})}+O(\frac{1}{P_{1}^{2}})\bigg\}.
\end{equation}
As we mentioned before the maximum efficiency always coming from
Carrnot cycle, here also we had maximum efficiency. So, one can
employed Carrnot efficiency at high pressure limit and for values of
$\Phi_{E}$ close to 1.

\section{Thermodynamical properties of dynamic  black hole and heat engine}
Now we investigate dynamic black hole system. One example of such a
system is $AdS$ rotating charged black hole solution which the
corresponding metric is given by [38],
\begin{equation}
ds^2=-\frac{\Delta}{\rho^2}\bigg[dt-\frac{a\sin^2\theta}{\Xi}d\varphi\bigg]^2+\frac{\rho^2}{\Delta}dr^2+\frac{\rho^2}{S}d\theta^2+S\frac{\sin^2\theta}{\rho^2}\bigg[adt-\frac{r^2+a^2}{\Xi}d\varphi\bigg]^2~,
\end{equation}
where in $d=4$ we have following expressions,
\begin{equation}
\rho^2=r^2+a^2\csc^2\theta~~,~~ \Xi=1-\frac{a^2}{\ell^2}~~,~~
\Delta=(r^2+a^2)(1+\frac{r^2}{\ell^2})-2mr+q^{2}~~,~~s=1-\frac{a^2}{\ell^2}\csc^2\theta~.
\end{equation}
The entropy, temperature, electric potential and angular velocity
are given by,
\begin{equation}
S=\pi\frac{(r_+^2+a^2)}{\Xi}~~,~~ T=\frac
{r_+(1+\frac{a^2}{\ell^2}+\frac{3r_+^2}{\ell^2}-\frac{a^2+q^2}{r_+^2})}{4\pi(r_+^2+a^2)}~~,~~\Phi=\frac{qr_{+}}{r_{+}^{2}+a^{2}}~~,~~
\Omega_H=\frac{a\Xi}{r_+^2+a^2}~,
\end{equation}
one can rewrite the temperature as follows,
\begin{equation}
T=\frac{r_+}{2S}-\frac{1}{4\pi r_{+}}+2Pr_{+}-\frac{q^{2}}{4\Xi
Sr_{+}}~.
\end{equation}
The mass of black hole $M$, the charge $Q$ and the angular momentum
$J$ are related to parameters $m$, $q$ and $a$ as follows,
\begin{equation}
M=\frac{m}{\Xi^2}  ~~~~,~~~~Q=\frac{q}{\Xi} ~~~~,~~~~
J=\frac{am}{\Xi^2}.
\end{equation}
The pressure is identified by,
$P=-\frac{\Lambda}{8\pi}=\frac{3}{8\pi}\frac{1}{\ell^2}$ and the
thermodynamic volume is [15,17],
\begin{equation}
V=\frac{2\pi}{3}\frac{(r_+^2+a^2)(2r_+^2\ell^2+a^2\ell^2-r_+^2a^2)+\ell^{2}q^{2}a^{2}}{\ell^2\Xi^2r_+},
\end{equation}
The equation of state is written  by following expression,
\begin{equation}
P=\frac{T}{\upsilon}-\frac{1}{2\pi \upsilon^2}+\frac{2Q^{2}}{\pi
\upsilon^{4}}+\frac{48J^2}{\pi
\upsilon^6}-\frac{96Q^{2}(24Q^{2}+5\upsilon^{2}+6\pi T
\upsilon^{3})J^{2}}{\pi \upsilon^6(8Q^{2}+\upsilon^{2}+\pi T
\upsilon^{3})^2}~,
\end{equation}
where
\begin{equation}
\upsilon=2\bigg(\frac{3V}{4\pi}\bigg)^\frac{1}{3}=2r_+
+\frac{12(3r_+^2+8\pi r_+^4P+Q^{2})J^2}{r_+(3r_+^2+8\pi
r_+^4P+3Q^{2})^2}.
\end{equation}
Now we are going to investigate the thermodynamical properties for
dynamic black hole (rotating black hole) with electric charge and
without charge.  First of all we consider zero charge as $Q=0$. ln
that case the equation of state is expressed as follows,
\begin{equation}
P=\frac{T}{\upsilon}-\frac{1}{2\pi \upsilon^2}+\frac{48J^2}{\pi
\upsilon^6}~,
\end{equation}
where
\begin{equation}
\upsilon=2\bigg(\frac{3V}{4\pi}\bigg)^\frac{1}{3}=2r_+
+\frac{12J^2}{r_+(3r_+^2+8\pi r_+^4P)}.
\end{equation}
In figure (4), we draw pressure with respect to specific volume. In
this figure, we have some phase transition and one can easily define
the holography heat engine for dynamic black hole (rotating black
hole). So, this property lead us to investigate corresponding black
hole as Carnot cycle and obtain the efficiency parameter. For this
reason, we have to prepare some thermodynamic quantities as heat
capacity, useful work and heat $Q_H$ to arrange corresponding
efficiency.

\begin{figure}
\hspace*{1cm}
\begin{center}
\epsfig{file=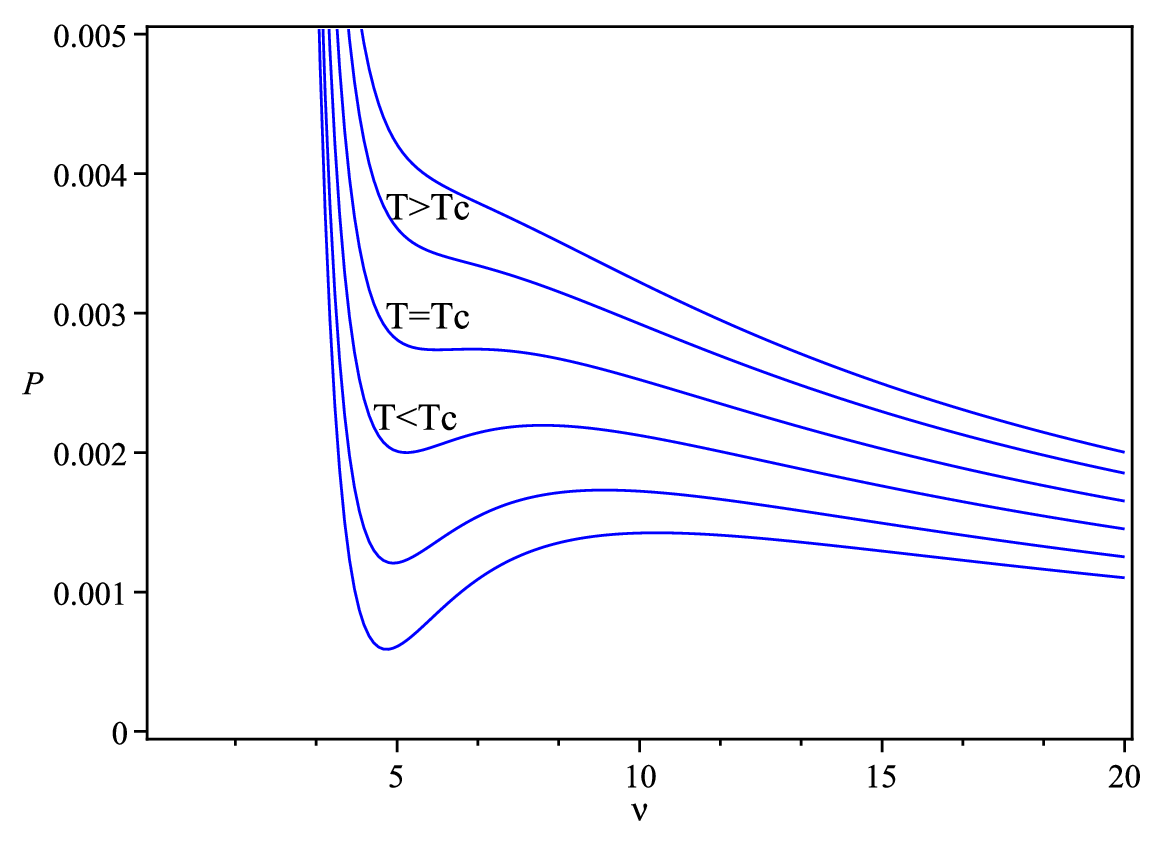,width=7cm} \caption{\small{$P-\upsilon$ diagram
for the rotating black hole for fixed J=1 and varying T.}}
\end{center}
\end{figure}
First of all we are going to calculate the heat capacity. The heat
capacity at constant pressure is given by,
\begin{equation}
C_P=\frac{2(\Xi S-a^2)}{\Xi}\bigg(\frac{\frac{1}{2S}-\frac{1}{4(\Xi
S-a^2)}+2P}{\frac{1}{2S}+\frac{1}{4(\Xi S-a^2)}+2P-\frac{(\Xi
S-a^2)}{S\Xi^2}}\bigg)~.
\end{equation}

By assuming $\frac{\pi a^{2}}{S\Xi}\ll1$, we can rewrite
thermodynamic volume as follows,
\begin{equation}
V=\frac{2S^{\frac{3}{2}}}{3\sqrt{\pi\Xi}}\bigg((1+\Xi)+\frac{\pi
a^{2}}{2S\Xi}(1-\Xi)\bigg),
\end{equation}
As we see, the volume is a function of $S$, $a$ and $\ell$. Then one
can say, these parameters will be constant at constant volume. On
the other hand, the temperature is dependent on $S$, $a$ and $\ell$.
So, $\frac{\partial T}{\partial S}$ is zero at constant volume
$(C_V=0)$. Therefore we define a cycle  as figure (5). The work done
along the isobars is,
\begin{equation}
W=\bigg( \frac{2(1+\Xi)}{3\sqrt{\pi\Xi}}\bigg(
S_{2}^{\frac{3}{2}}-S_{1}^{\frac{3}{2}}\bigg)+\frac{a^{2}}{3}\sqrt{\frac{\pi}{\Xi^{3}}}(1-\Xi)\bigg(
S_{2}^{\frac{1}{2}}-S_{1}^{\frac{1}{2}}\bigg)\bigg)\bigg(P_{1}-P_{4}\bigg).
\end{equation}
We take the high temperature and pressure limit to have explicit
expressions for the efficiency. Therefor, by expanding entropy and
heat capacity at constant pressure and large $T$, we can obtain the
following expressions,
\begin{eqnarray}
S=\frac{\pi T^{2}}{4\Xi P^{2}}-\frac{(2\Xi-1)}{4\Xi
P}-\frac{(2\Xi-1)^{2}}{16\pi \Xi T^{2}}+...\nonumber
\\&& \hspace{-80mm}
C_P=\frac{\pi T^{2}}{2\Xi P^{2}}-\frac{(2\Xi-1)}{2\Xi
P}-\frac{(2\Xi-1)^{2}}{8\pi \Xi T^{2}}+...~,
\end{eqnarray}
\begin{figure}
\hspace*{1cm}
\begin{center}
\epsfig{file=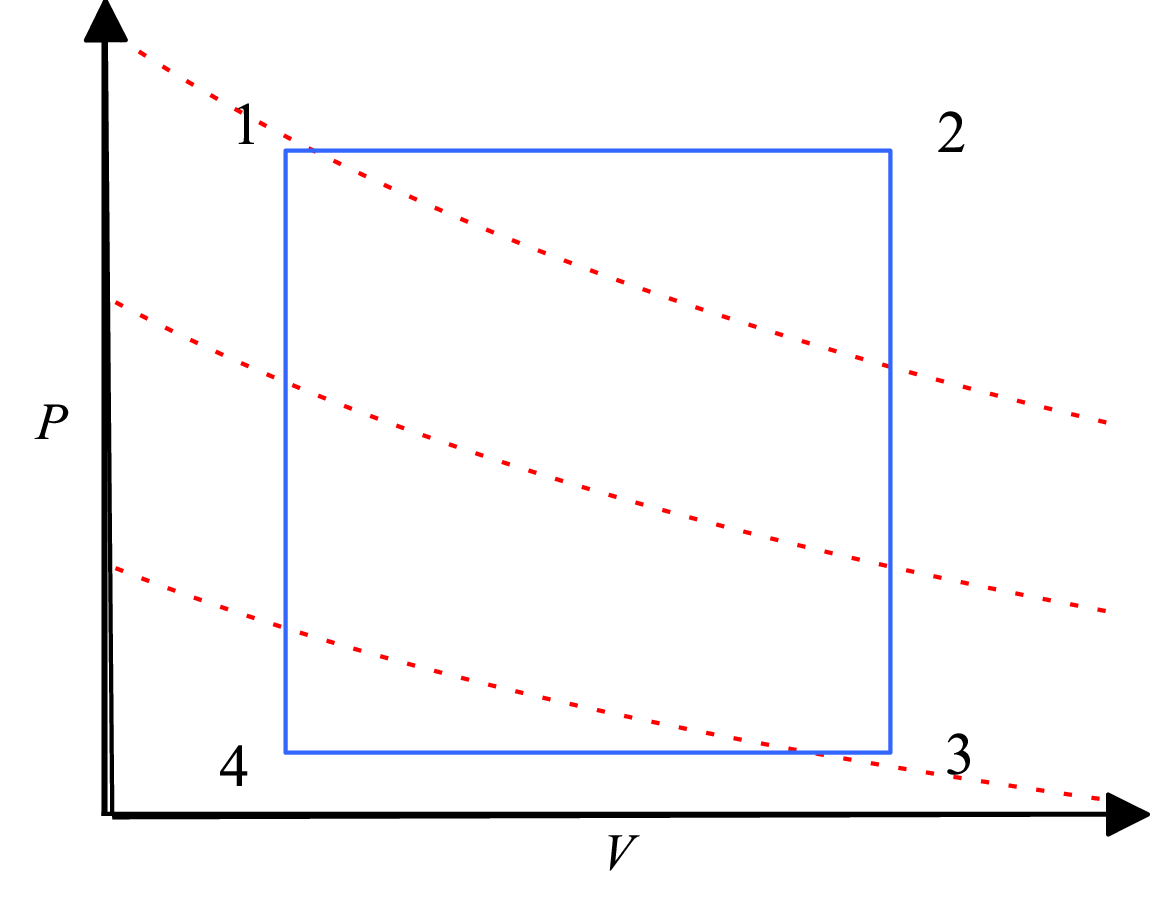,width=7cm} \caption{\small{Our other engine.}}
\end{center}
\end{figure}
Consequently, the heat $Q_H$ is as follows,
\begin{equation}
Q_H=\frac{4P_{1}}{3}\sqrt{\frac{\Xi}{\pi}}\bigg(
S_{2}^{\frac{3}{2}}-S_{1}^{\frac{3}{2}}\bigg)-\frac{(2\Xi-1)}{\sqrt{\pi\Xi}}\bigg(
S_{2}^{\frac{1}{2}}-S_{1}^{\frac{1}{2}}\bigg)+\frac{(2\Xi-1)^{2}}{16P_{1}\sqrt{\pi\Xi^{3}}}\bigg(
S_{2}^{-\frac{1}{2}}-S_{1}^{-\frac{1}{2}}\bigg)\bigg).
\end{equation}
Thus, the efficiency of this cycle is,
\begin{equation}
\eta=(1 - \frac{T_C}{T_H})(\frac{1+\Xi}{2\Xi})(1-\beta),
\end{equation}
where
\begin{equation}
\beta=\frac{\pi
a^{2}(\Xi-1)}{2\Xi(\Xi+1)}\frac{(S_{2}^{\frac{1}{2}}-S_{1}^{\frac{1}{2}})}{(S_{2}^{\frac{3}{2}}-S_{1}^{\frac{3}{2}})}+\frac{3(1-2\Xi)}{4\Xi
P_{1}}\frac{(S_{2}^{\frac{1}{2}}-S_{1}^{\frac{1}{2}})}{(S_{2}^{\frac{3}{2}}-S_{1}^{\frac{3}{2}})}+...~.
\end{equation}

From Eqs. (44) and (45), it is obvious that the defined cycle has
the Carrnot efficiency for very small values of rotating parameter
($a\ll1$). Figure (6) show the ratio $\frac{\eta}{\eta_{C}}$ with
respect to $a$ for three values of pressure ($P_{1}=1$, $P_{1}=2$
and $P_{1}=5$). We noted that, the corresponding black hole has not
physical solution for $P_{1}<1$. And at $P_{1}=1$, the obtained
efficiency is larger than the Carrnot efficiency in range of
$0<a<0.48$ which is inconsistent with the second law of
thermodynamic. But $a>0.48$ satisfy the second law. At $P_{1}=2$ and
$P_{1}=5$, the obtained efficiency is consistent with the second law
for $a>0.34$ and $a>0.21$ respectively. It show that, the
corresponding engine has a physical efficiency for any value of
rotating parameter $a$ at large pressure limit. As we see for $a>1$,
the efficiency reaches a specific value
$(\eta_{max}\simeq\frac{1}{2}\eta_{C})$  which
this consequence is independent of pressure value.\\
Now we are going to investigate the heat engine of corresponding
solution in presence of electric charge as $(Q\neq 0).$
\begin{figure}
\hspace*{1cm}
\begin{center}
\epsfig{file=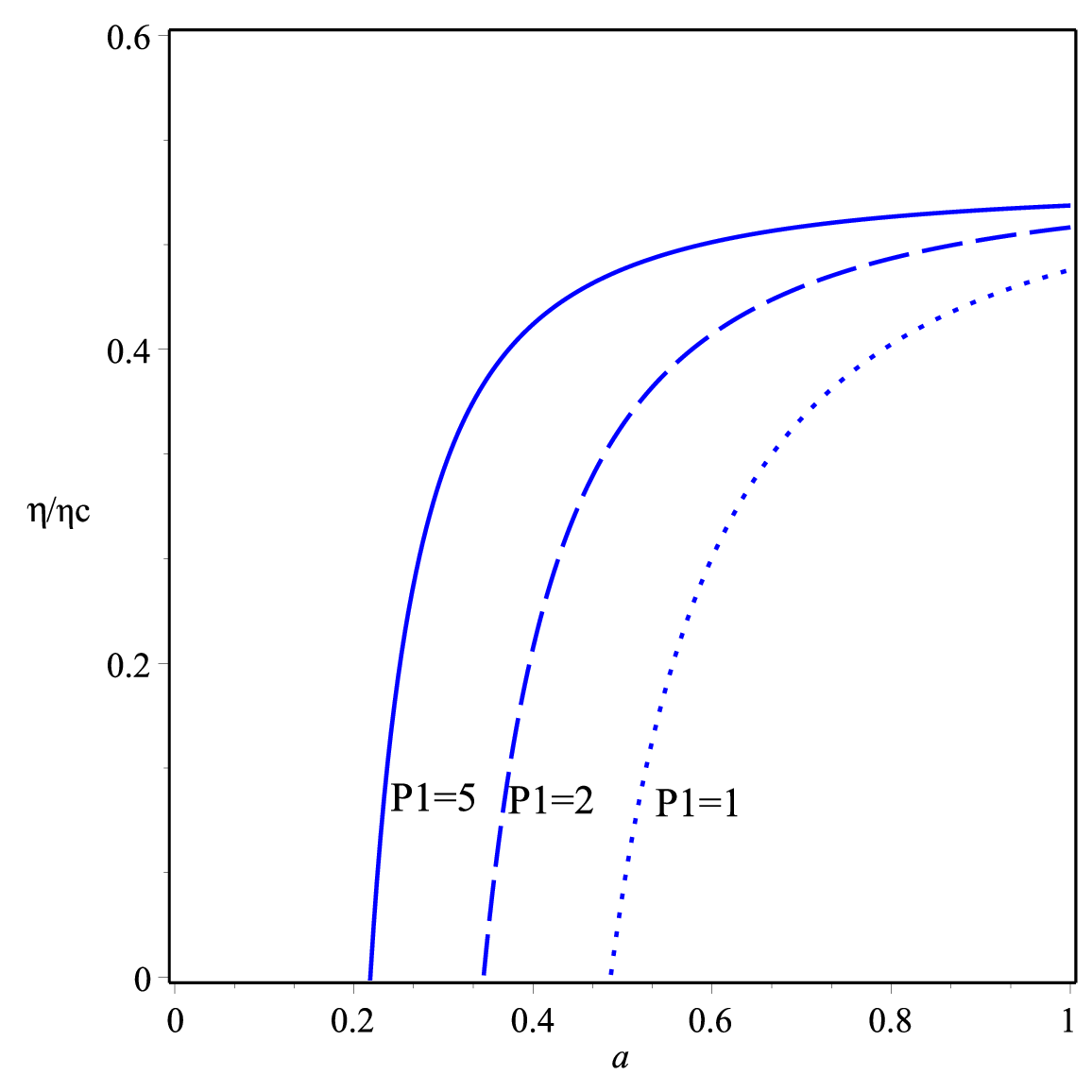,width=7cm} \caption{\small{The ratio
$\frac{\eta}{\eta_{C}}$ with respect to $a$ for fixed ( $S_{2}=20$
and $S_{1}=10$ ) and varying $P_{1}$.}}
\end{center}
\end{figure}
By using equation (35), we can investigate phase structure of
rotating charged black hole. We draw pressure with respect to
specific volume in figure (7). As we see, the corresponding black
hole solution has a Van der Walss-like behavior. Therefore, we can
define the holography heat engine for this solution.
\begin{figure}
\hspace*{1cm}
\begin{center}
\epsfig{file=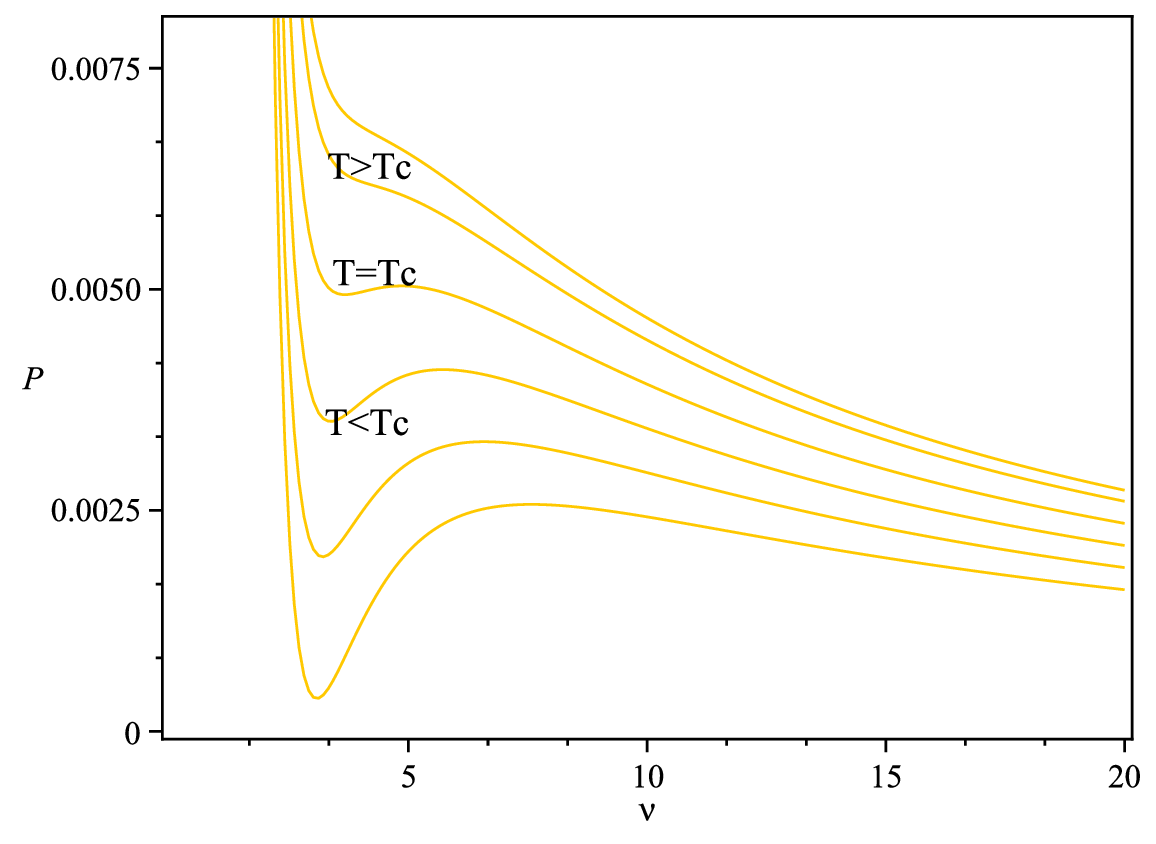,width=7cm} \caption{\small{$P-\upsilon$ diagram
for the rotating charged black hole for fixed q=0.5, J=0.4 and
varying T.}}
\end{center}
\end{figure}
First of all we calculate the heat capacity at constant pressure,
which is given by,
\begin{equation}
C_P=\frac{2(\Xi S-a^2)}{\Xi}\bigg(\frac{\frac{1}{2S}-\frac{1}{4(\Xi
S-a^2)}+2P-\frac{\pi q^2}{4\Xi S(\Xi
S-a^2)}}{\frac{1}{2S}+\frac{1}{4(\Xi S-a^2)}+2P+\frac{\pi q^2}{4\Xi
S(\Xi S-a^2)}+\frac{\pi q^2}{2\Xi^{2}S^2}-\frac{(\Xi
S-a^2)}{S\Xi^2}}\bigg)~.
\end{equation}
Here,  also we rewrite thermodynamic volume by assuming $\frac{\pi
a^{2}}{S\Xi}\ll1$,

\begin{equation}
V=\frac{2S^{\frac{3}{2}}}{3\sqrt{\pi\Xi}}\bigg((1+\Xi)+\frac{\pi
a^{2}}{2S\Xi}(1-\Xi)+\frac{\pi^{\frac{3}{2}}q^{2}a^{2}}{\Xi^{\frac{5}{2}}S^{\frac{1}{2}}}\bigg),
\end{equation}
Here also we will have $C_V=0$ similar to pervious case $(Q=0)$. In
that case,  we can define a cycle as figure (8) and obtain the work
done along the isobars,

\begin{equation}
W=\bigg( \frac{2(1+\Xi)}{3\sqrt{\pi\Xi}}\bigg(
S_{2}^{\frac{3}{2}}-S_{1}^{\frac{3}{2}}\bigg)+\frac{a^{2}}{3}\sqrt{\frac{\pi}{\Xi^{3}}}(1-\Xi)\bigg(
S_{2}^{\frac{1}{2}}-S_{1}^{\frac{1}{2}}\bigg)+\frac{2\pi^{\frac{3}{2}}q^{2}a^{2}}{3\Xi^{\frac{5}{2}}S^{\frac{1}{2}}}\bigg(
S_{2}^{-\frac{1}{2}}-S_{1}^{-\frac{1}{2}}\bigg)\bigg)\bigg(P_{1}-P_{4}\bigg).
\end{equation}
As before, we obtain $S$ and $C_P$ by expanding around large
pressure and temperature,
\begin{eqnarray}
S=\frac{\pi T^{2}}{4\Xi P^{2}}-\frac{(2\Xi-1)}{4\Xi
P}-\frac{(2\Xi-1)^{2}}{16\pi \Xi T^{2}}+...\nonumber
\\&& \hspace{-80mm}
C_P=\frac{\pi T^{2}}{2\Xi P^{2}}-\frac{(2\Xi-1)}{2\Xi
P}-\frac{(2\Xi-1)^{2}}{8\pi \Xi T^{2}}+...~,
\end{eqnarray}
\begin{figure}
\hspace*{1cm}
\begin{center}
\epsfig{file=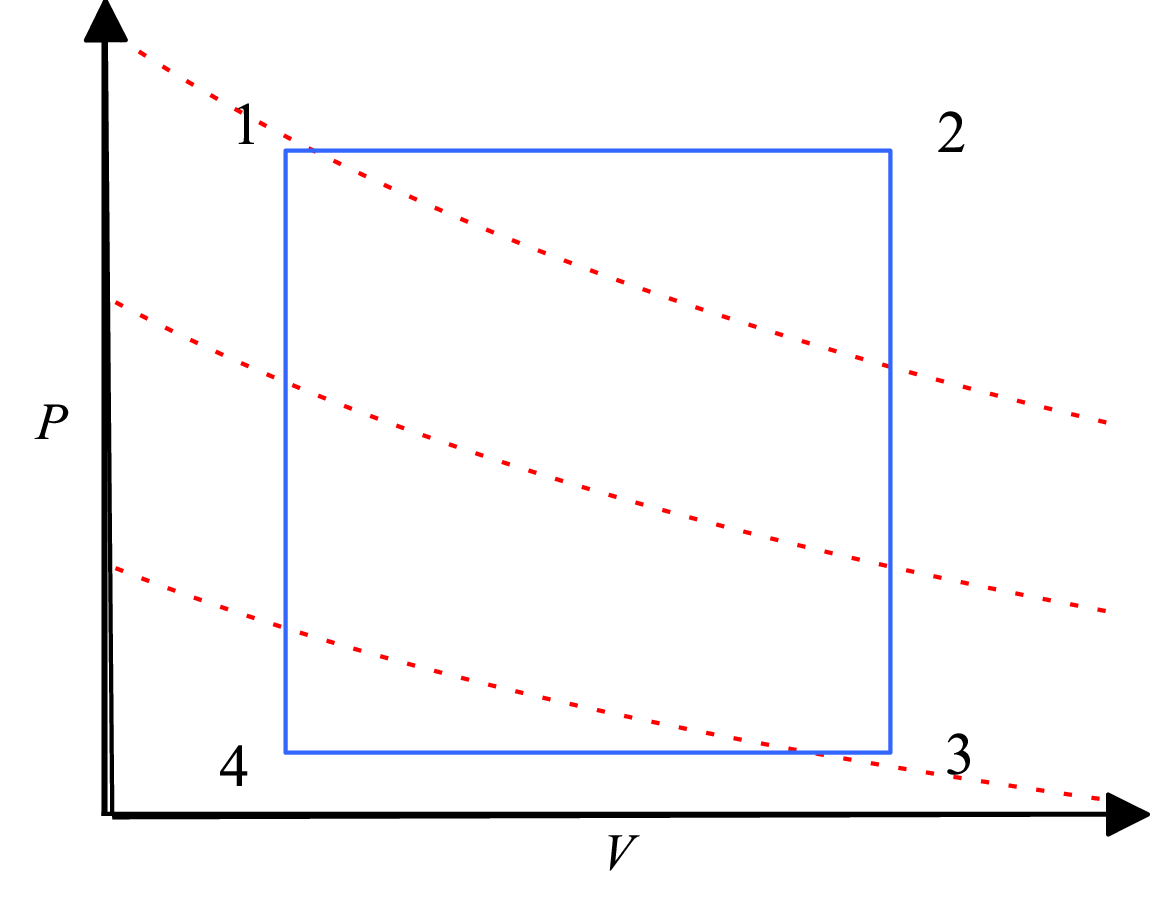,width=7cm} \caption{\small{Our other engine.}}
\end{center}
\end{figure}
Consequently, the heat $Q_H$ is as follows,
\begin{equation}
Q_H=\frac{4P_{1}}{3}\sqrt{\frac{\Xi}{\pi}}\bigg(
S_{2}^{\frac{3}{2}}-S_{1}^{\frac{3}{2}}\bigg)-\frac{(2\Xi-1)}{\sqrt{\pi\Xi}}\bigg(
S_{2}^{\frac{1}{2}}-S_{1}^{\frac{1}{2}}\bigg)+\frac{(2\Xi-1)^{2}}{16P_{1}\sqrt{\pi\Xi^{3}}}\bigg(
S_{2}^{-\frac{1}{2}}-S_{1}^{-\frac{1}{2}}\bigg)\bigg).
\end{equation}
Therefore, the efficiency of this cycle is,
\begin{equation}
\eta=(1 - \frac{T_C}{T_H})(\frac{1+\Xi}{2\Xi})(1-\beta),
\end{equation}
where
\begin{equation}
\beta=\frac{\pi
a^{2}(\Xi-1)}{2\Xi(\Xi+1)}\frac{(S_{2}^{\frac{1}{2}}-S_{1}^{\frac{1}{2}})}{(S_{2}^{\frac{3}{2}}-S_{1}^{\frac{3}{2}})}+\frac{3(1-2\Xi)}{4\Xi
P_{1}}\frac{(S_{2}^{\frac{1}{2}}-S_{1}^{\frac{1}{2}})}{(S_{2}^{\frac{3}{2}}-S_{1}^{\frac{3}{2}})}-\frac{\pi^{2}q^{2}a^{2}}{\Xi^{2}(\Xi+1)}\frac{(S_{2}^{-\frac{1}{2}}-S_{1}^{-\frac{1}{2}})}{(S_{2}^{\frac{3}{2}}-S_{1}^{\frac{3}{2}})}+...~.
\end{equation}

\begin{figure}
\hspace*{1cm}
\begin{center}
\epsfig{file=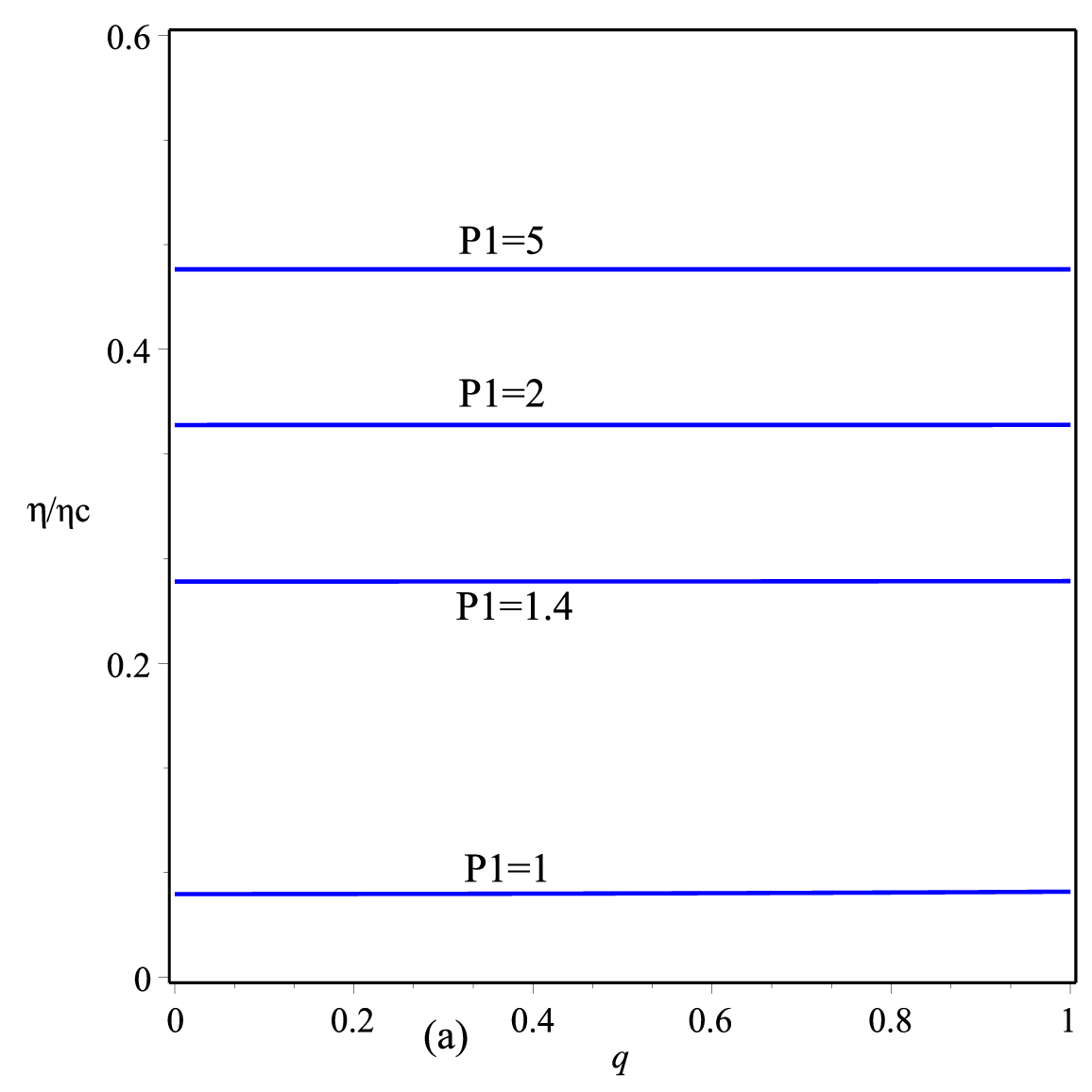,width=7cm}
\epsfig{file=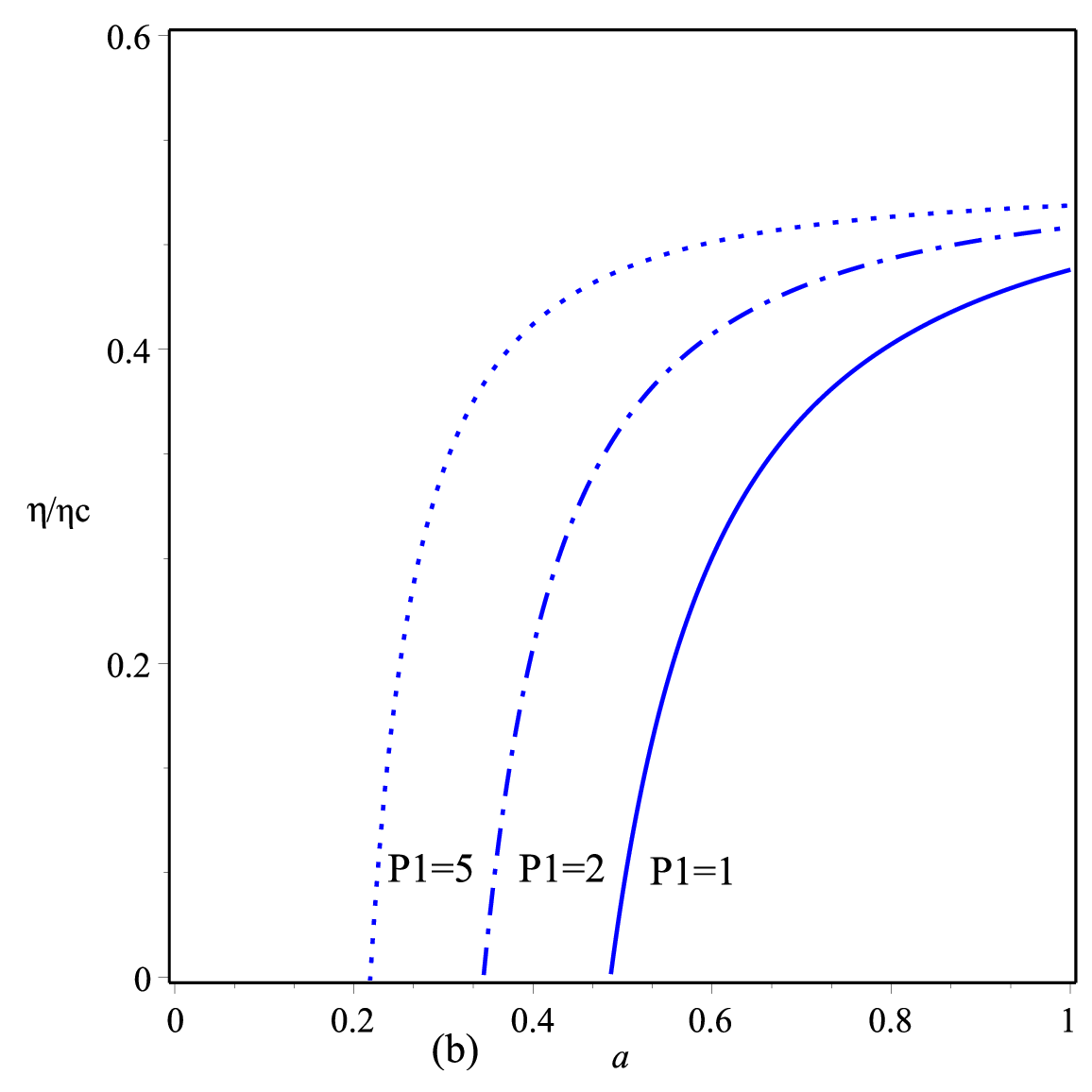,width=7cm}\caption{\small{(a) The ratio
$\frac{\eta}{\eta_{C}}$ with respect to $q$ for fixed ($a=0.5$,
$S_{2}=20$ and $S_{1}=10$ ) and varying $P_{1}$. (b) The ratio
$\frac{\eta}{\eta_{C}}$ with respect to $a$ for fixed ($q=0.5$,
$S_{2}=20$ and $S_{1}=10$ ) and varying $P_{1}$.}}
\end{center}
\end{figure}
The obtained efficiency is same to equation (44) and the only
difference is the term of electric charge. In figure (9a), we
considered the rotating parameter as a constant and draw the ratio
$\frac{\eta}{\eta_{C}}$ with respect to $q$. As we see, the
efficiency increases by increasing pressure and it doesn't change by
varying the charge. Then we include that the charge of black hole
has any role in efficiency of this black hole solution. We also draw
the ratio $\frac{\eta}{\eta_{C}}$ with respect to $a$ for fixed $q$
in figure (9b). The figure is exactly identical to figure (6). By
comparing figures (9a) and (9b), we noted that the rotating
parameter $a$ has a more effective role than $q$ in the
corresponding efficiency.

\section{Conclusion}
In this paper we have studied thermodynamic cycle and heat engine
for different black holes. We considered the holographic heat engine
for a static black hole (Dyonic BH) in constant electric potential
and magnetic charge ensemble. We noticed that at high pressure and
temperature limit, the corresponding heat engine has a maximum
efficiency. Then we continued our investigation and considered a
dynamic black hole (rotating charged black hole). For first case
(the solution without electric charge), We saw that, the physical
solution existed for only $P_{1}>1$. And at large pressure limit the
corresponding engine had a physical efficiency for any value of
rotating parameter. Also we noted that, the corresponding efficiency
will have maximum value $(\eta_{max}\simeq\frac{1}{2}\eta_{C})$ for
$a>1$. Then we considered a rotating black hole along with electric
charge and defined a cycle for the corresponding solution. We saw
the obtained result of rotating charged black hole is identical to
rotating black hole. In other words the efficiency is independent of
charge value and only the rotating parameter $a$ has a key role in
corresponding efficiency.

\end{document}